\documentclass[fleqn,usenatbib]{mnras}

\usepackage[left]{lineno}
\usepackage{epstopdf}
\usepackage{ulem}
\usepackage{amssymb}
\usepackage{times}
\usepackage[usenames,dvipsnames]{pstricks}
\usepackage{psfrag}
\usepackage{lipsum}
\usepackage{natbib}
\newcommand{\be}{\begin{equation}}
\newcommand{\ee}{\end{equation}}
\newcommand{\bary}{\begin{eqnarray}}
\newcommand{\eary}{\end{eqnarray}}

\usepackage{newtxtext,newtxmath}

\usepackage[T1]{fontenc}

\DeclareRobustCommand{\VAN}[3]{#2}
\let\VANthebibliography\thebibliography
\def\thebibliography{\DeclareRobustCommand{\VAN}[3]{##3}\VANthebibliography}

\usepackage{graphicx}	
\usepackage{amsmath}	

\title[Early afterglow of GRB 190829A]{The early afterglow of GRB 190829A}

\author[S.~Dichiara et al.]{
S.~Dichiara,$^{1,2,3}$\thanks{E-mail: sbd5667@psu.edu}
E.~Troja,$^{1,2}$
V.~Lipunov,$^{4}$
R.~Ricci,$^{5,6}$
S.~R.~Oates,$^{7}$
N.~R.~Butler,$^{8}$
E.~Liuzzo,$^{6}$
G.~Ryan,$^{1,2,9}$\newauthor
B.~O'Connor,$^{1,2,10,11}$
S.~B.~Cenko,$^{2,12}$
R.~G.~Cosentino,$^{13}$
A.~Y.~Lien,$^{14}$
E.~Gorbovskoy,$^{4}$
N.~Tyurina,$^{4}$
P.~Balanutsa,$^{4}$\newauthor
D.~Vlasenko,$^{4}$
I.~Gorbunov,$^{4}$
R.~Podesta,$^{15}$
F.~Podesta,$^{15}$
R.~Rebolo,$^{16}$
M.~Serra,$^{16}$
D.~A.~H.~Buckley,$^{17}$
\\
$^{1}$ Department of Astronomy, University of Maryland, College Park, MD 20742-4111, USA \\
$^{2}$ Astrophysics Science Division, NASA Goddard Space Flight Center, 8800 Greenbelt Rd, Greenbelt, MD 20771, USA \\
$^{3}$ Department of Astronomy and Astrophysics, The Pennsylvania State University, 525 Davey Lab, University Park, PA 16802, USA \\
$^{4}$ Lomonosov Moscow State University, SAI, Physics Department, 119234,
Universitetskiy pr., 13, Moscow,Russia \\
$^{5}$ Istituto Nazionale di Ricerche Metrologiche - Torino, Italy \\
$^{6}$ INAF - Istituto di Radioastronomia - Bologna, Italy \\
$^{7}$ School of Physics and Astronomy \& Institute for Gravitational Wave Astronomy, University of Birmingham, B15 2TT, UK\\
$^{8}$ School of Earth and Space Exploration, Arizona State University, Tempe, AZ 85287, USA\\
$^{9}$ Perimeter Institute for Theoretical Physics, 31 Caroline St. N., Waterloo, ON, N2L 2Y5, Canada\\
$^{10}$ Department of Physics, The George Washington University, 725 21st Street NW, Washington, DC 20052, USA\\
$^{11}$ Astronomy, Physics and Statistics Institute of Sciences (APSIS), The George Washington University, Washington, DC 20052, USA\\
$^{12}$ Joint Space-Science Institute, University of Maryland, College Park, MD 20742 USA\\
$^{13}$ Space Telescope Science Institute, 3700 San Martin Drive, Baltimore, MD 21218, USA\\
$^{14}$ University of Tampa, Department of Chemistry, Biochemistry, and Physics, 401 W. Kennedy Blvd, Tampa, FL 33606, USA\\
$^{15}$ Observatorio Astronomico Felix Aguilar(OAFA), Avda Benavides s/n, Rivadavia, El Leonsito, Argentina\\
$^{16}$ Instituto de Astrofisica de Canarias, Via Lactea, E-38200 La Laguna,
Tenerife, Spain\\
$^{17}$ South African Astronomical Observatory, PO Box 9, Observatory Road, Observatory 7935, Cape Town, South Africa
}

\date{Accepted XXX. Received YYY; in original form ZZZ}

\pubyear{2015}

\begin{document}
\label{firstpage}
\pagerange{\pageref{firstpage}--\pageref{lastpage}}
\maketitle

\begin{abstract}

GRB 190829A  at $z$=0.0785 is the fourth closest long GRB ever detected by the {\it Neil Gehrels Swift observatory}, and the third confirmed case with a very high energy component. 
We present our multi-wavelength analysis of this rare event, focusing on its early stages of evolution, and including data from {\it Swift}, the MASTER global network of optical telescopes, ALMA, and ATCA. 
We report sensitive limits on the linear polarization of the optical emission, 
disfavouring models of off-axis jets to explain the delayed afterglow peak.
The study of the multi-wavelength light curves and broadband spectra 
supports a model with at least two emission components: 
a bright reverse shock emission, visible at early times in the optical and X-rays and, later, in the radio band; and a forward shock component dominating at later times and lower radio frequencies. 
A combined study of the prompt and afterglow properties shows many similarities with cosmological long GRBs, suggesting that GRB~190829A is an example of classical GRBs in the nearby universe. 
 
\end{abstract}

\begin{keywords}
gamma-ray burst: general -- gamma-ray burst: individual (GRB190829A) -- radiation mechanisms: non-thermal
\end{keywords}
\section{Introduction}

The discovery of a low-redshift ($z \lesssim$0.1) gamma-ray burst (GRB) is not a frequent occurrence, yet it provides us with a rare opportunity to study in detail the properties of these extreme explosions, their progenitors and local environment  \citep[e.g.][]{Galama1998,Mazzali01,Kouveliotou04,Kruhler17,Arabsalmani20}. 
The GRB proximity allows for a close-up view that can not be achieved with the study of the larger sample of events at cosmological distances. 
Furthermore, these nearby bursts appear to probe a population of sub-energetic explosions powered by mildly relativistic ejecta, allowing us to  fill the observational gap between ordinary supernovae (SNe) and relativistic GRBs \citep{Soderberg04Nature,Soderberg06,Corsi17,Ho20}. 

In over 15 years of operations of NASA's {\it Swift} mission \citep{Gehrels04}, the long duration GRB~190829A located at a redshift $z$=0.0785 is the fourth closest event ever observed, after GRB 060218 \citep[][$z$=0.0331]{Campana2006,Soderberg06}, GRB 100316D \citep[][$z$=0.059]{Starling2011} and GRB 171205A \citep[][$z$=0.0368]{Izzo2019}. 
As such, it is characterized
by an exquisite multi-wavelength dataset spanning over fifteen decades in energy \citep{Rhodes2020,HESS2021,Salafia2021}.
Similarly to other nearby long GRBs \citep[e.g.][]{WoosleyBloom06,Cano17}, it was followed by a bright broad-lined Type Ic supernova SN~2019oyw \citep{Hu2021}.

\begin{figure*}
\centering
\includegraphics[scale=0.38, clip]{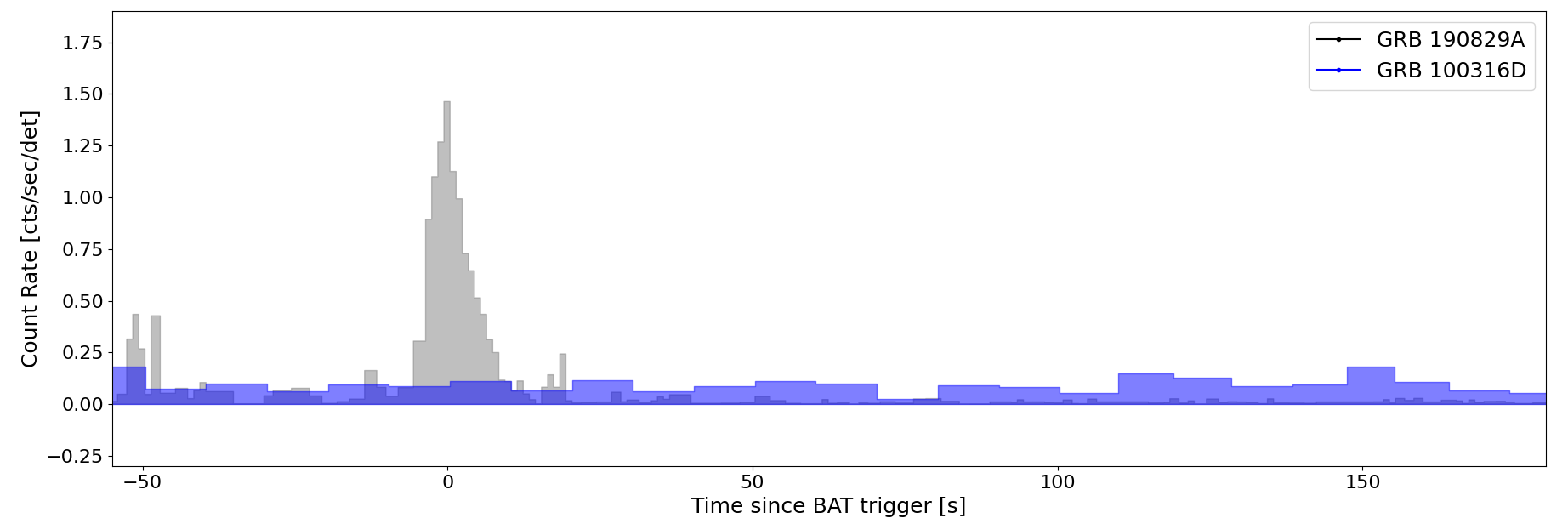}
\caption{{\it Swift}/BAT mask-weighted light curve of GRB 190829A in the energy range 15--350 keV with a time bin of 1~s. A weak precursor is visible at around 50~s before the trigger. For comparison we show the BAT light curve of GRB~100316D with a time bin of 10~s and rescaled by a factor of 10 for plotting purposes. The two nearby bursts show markedly different temporal properties. 
\label{fig:LCs}}
\end{figure*}

One of the most remarkable features of this GRB is the detection
of very high energy (VHE; $>$ 100 GeV) emission by the imaging atmospheric Cherenkov telescope High Energy Stereoscopic System (H.E.S.S.; \citealt{HESS2021}),
which makes this only the third example of a long GRB with a VHE counterpart after 190114C and 180720B \citep{MAGIC2019,HESS2019}.
The VHE component of GRB~190829A displays a soft spectrum with photon index
$\Gamma$\,$\approx$2 and decays in time as $t^{-1.1}$ between 4 and 56 hours after the burst. 

Broadband observations, from radio to X-rays, of the early afterglow are key to understand the origin of the VHE component.  
For instance, based on the similar temporal and spectral properties of the X-ray and VHE emission, \citealt{HESS2021} argued that they arise from the same spectral component. 
However, the interpretation of the VHE photons as non-thermal synchrotron radiation is problematic as it pushes the boundary of the maximum allowed energy 
\citep{pirannakar10,Zou11,Sironi13}. 

\citet{Sato2021} discussed instead an off-axis two-component jet model to explain the peculiar properties of this event, such as its low gamma-ray energy and delayed afterglow peak \citep[e.g.][]{Chand20,Fraija2020}. 
In this model, the early afterglow, including the VHE component, is powered by a narrow relativistic jet seen off-axis. A second wider and slower jet, seen close to its axis, is instead responsible for the late-time afterglow emission. 
An alternative model was suggested by \citet{Rhodes2020} and \citet{Salafia2021}, who discussed a forward shock plus reverse shock scenario to interpret the complex afterglow evolution, from radio to X-rays. 
\citet{Salafia2021} further showed that the low energy and VHE emission can be self-consistently described as 
synchrotron and synchrotron Self-Compton radiation, respectively. However, within this model, the kinetic energy of the explosion is rather high ($E_{K,\rm iso}$\,$\approx$\,$2\times 10^{53}$ erg),
and implies an unusually low ($\eta_{\gamma}\lesssim$1\%) radiative efficiency. 

In this work, we present the multi-wavelength {\it Swift} observations of GRB~190829A, 
and complement them with optical data from the MASTER telescopes, and radio data from ATCA and ALMA to fully characterize the early stages of the afterglow evolution. 
The paper is organized as follows: in Section~\ref{sec:data} we present the data reduction and analysis of prompt and afterglow observations. Results are presented in Section~\ref{sec:results} where we discuss the prompt emission properties and the multi-wavelength evolution of the early afterglow, discussing their implications in terms of emission mechanism, physics of the ejecta and properties of the external medium.  
Conclusions are summarized in Section~\ref{sec:conclusions}. Uncertainties are quoted at the 1$\sigma$ confidence level for each parameter of interest and upper limits are given at a 2 $\sigma$ level, unless stated otherwise. We adopted a standard $\Lambda$CDM cosmology \citep{2018arXiv180706209P}.

\section{Data Analysis}\label{sec:data}

\subsection{{\it Swift}/BAT}

The Burst Alert Telescope \citep[BAT,][]{barthelmy2005}
on board {\it Swift} instrument triggered on GRB 190829A at $T_0$=19:56:44.60 UTC of August 8, 2019 \citep{2019GCN.25552....1D}. 
BAT data were processed using \textsc{HEASOFT} package (v6.25). The energy calibration was applied with \textsc{bateconvert} and the mask weighting was included with \textsc{batmaskwtevt}.
The partial coding for this event was only 4.2\%.
We used \textsc{battblocks} to run the Bayesian Block algorithm over the 64 ms, background-subtracted, 15--350 keV light curve finding a $T_{90} = 57 \pm 3$~s and $T_{50} = 5 \pm 1$~s \footnote{$T_{90}$ ($T_{50}$) is the time during which the cumulative time counts increase from 5 to 95\% (from 25 to 75\%) above background \citep{1993ApJ...413L.101K}}. 
\textsc{battblocks} was run with the default configuration options except the background-subtraction parameter \textit{bkgsub} set on `YES'. Spectral analysis was carried out using \textsc{XSPECv12.10.1} \citep{Arnaud1996}

The BAT light curve shows two main episodes of emission (Figure~\ref{fig:LCs}). 
A first peak (precursor) occurs about $T_0$-50~s before the trigger and lasts about 5~s. 
Its spectrum, integrated from $T_0$-52~s to $T_0$-47~s, is well fit by a simple power law with a photon index of $1.89 \pm 0.22$ ($\chi^{2}/\mathrm{d.o.f.} = 1.04$) and a fluence of $f_{\gamma} = (8.8 \pm 1.3) \times 10^{-7}~\mathrm{erg\,cm^{-2}}$ (15--150 keV). 
The main pulse starts at $T_0$-4~s, peaks at $T_0$-0.5~s, and lasts about 13~s. Its spectrum is well described  by a simple power law with a photon index $\Gamma$=$2.59 \pm 0.09$ ($\chi^{2}/\mathrm{d.o.f.} = 1.06$), significantly softer
than the precursor emission. 
During this second episode, the fluence is $f_{\gamma} = (5.76 \pm 0.10) \times 10^{-6}~\mathrm{erg\,cm^{-2}}$ (15--150 keV) and the hardness ratio is S(25-50)/S(50-100)=1.49 $\pm$ 0.20. 

\subsection{{\it Swift}/XRT}

The {\it Swift} X-ray Telescope \citep[XRT;][]{burrows2005}
rapidly slewed to the GRB position. Observations started 87.3~s  after the BAT trigger, collecting data in Windowed Timing (WT) while the spacecraft was slewing to the burst location.
The astrometrically corrected X-ray position \citep{evens2007} was derived using the alignment between XRT and the UltraViolet and Optical Telescope \citep[UVOT;][]{Roming2005} and matching UVOT field sources to the USNO-B1 catalog.
 Its value is $\alpha$ = 02$^{h}$ 58$^{m}$ 10.57$^{s}$, $\delta$ = -08$^{\circ}$ 57$\arcmin$ 30.1$\arcsec$ (J2000.0) with an estimated uncertainty of 1.8$\arcsec$ (radius, 90\% confidence including systematic error).
 
\begin{figure*}
\centering
\includegraphics[scale=0.40, clip]{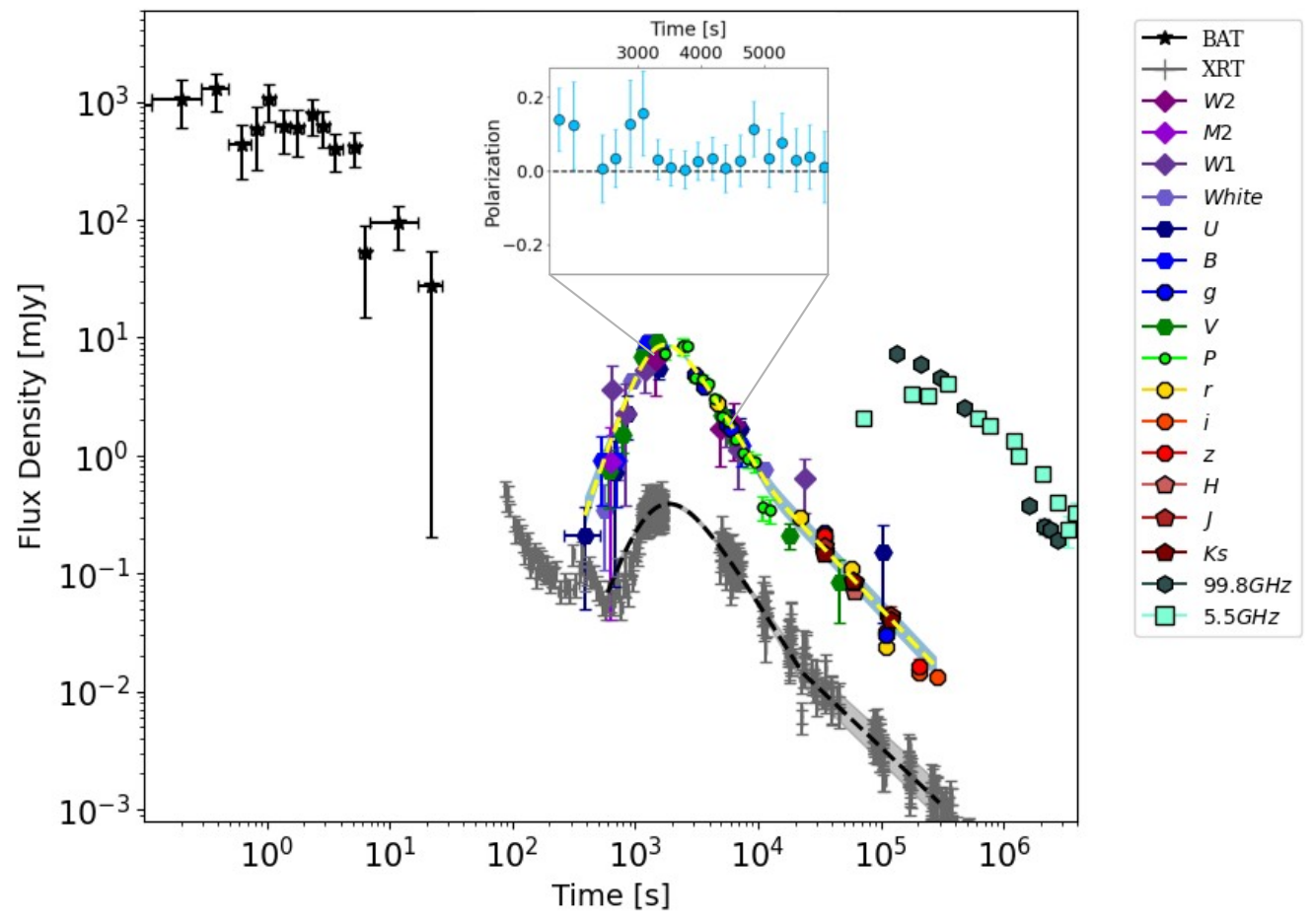}
\hspace*{-3cm}
\caption{Multiwavelength light curve of GRB 190829A. BAT and XRT unabsorbed flux densities are derived at 1 keV. 
 Optical points are re-normalized to the UVOT $v$-band and then scaled by a factor of 5 for plotting purposes. The best fit temporal model of the optical and X-ray data is indicated by the dashed yellow and dashed black line, respectively.  
 The insert shows the absolute values of the Stokes parameter Q from MASTER polarization measurements. 
\label{multiwave_lc}}
\end{figure*}

  The X-ray spectrum\footnote{https://www.swift.ac.uk/xrt\_spectra/00922968/}, 
  integrated between 235~s to 46~ks, is well fit with an absorbed power-law with photon index of 2.19$\pm$0.08 and an intrinsic absorption of $N_H=(1.4\pm 0.10)\times 10^{22}$ cm$^{-2}$
  in addition to the Galactic value $N_H=5.6\times 10^{20}$ cm$^{-2}$  \citep{Willingale2013}.
  The count rate light curve integrated over the 0.3--10 keV energy band was retrieved from the online repository\footnote{https://www.swift.ac.uk/xrt\_curves/00922968/} and converted to unabsorbed energy flux using a conversion factor of 
  1.20 $\times$ 10$^{-10}$ erg cm$^{-2}$ ct$^{-1}$ derived from the best fit spectral parameters.
  We used the measured photon index of 2.19 to convert fluxes into flux densities, as presented in Figure~\ref{multiwave_lc}.

\subsection{{\it Swift}/UVOT}
{\it Swift}/UVOT
began settled observations of the field of GRB 190829A 106 s after 
the trigger \citep{2019GCN.25552....1D}. The afterglow is detected in $white$, $v$, $b$, $u$ and $uvw1$, 
but not in the $uvm2$ and $uvw2$ filters. 
The analysis pipeline used software HEADAS 6.24 and UVOT calibration 20170922.
The source counts were extracted using a region of 3\arcsec radius. In order to be consistent with the UVOT calibration, these count rates were then corrected to 5\arcsec using the curve of growth contained in the calibration files. Background counts were extracted using a circular region of radius 20" from a blank area 
of sky situated near to the source position. The count rates were obtained from the image lists using the {\it Swift} tools \textsc{uvotsource}.

The GRB lies within a bright nearby galaxy, identified as 
2MASX J02581029-0857189 at $z$ = 0.0785 \cite{2019GCN.25565....1V},
which significantly contribute to the measured UVOT count rates.
In order to estimate the level of contamination, 
for each filter we combined all the late time exposures, from $58$ days 
until the end of observations ($\sim$122 days after the trigger).
At these late times we can assume that the emission from the afterglow and supernova is negligible compared to the host after this time (both components have AB magnitude $>$23 at this stage). 
We extracted the count rate in the late combined exposures using the same 3\arcsec radius and aperture corrected this to 5\arcsec. 
We derive the following values for the host galaxy's contribution: 1.14 $\pm$ 0.05 ct/s in $white$, 0.26 $\pm$ 0.02 ct/s in $v$, 0.33 $\pm$ 0.02 ct/s in $b$, 0.15 $\pm$ 0.01 ct/s in $u$, 0.04 $\pm$ 0.02 ct/s in $uvw1$, 0.03 $\pm$ 0.01 ct/s in $uvm2$ and 0.04 $\pm$ 0.02 ct/s in $uvw2$.
These were subtracted from the source count rates to obtain the afterglow count rates. These host subtracted count rate were converted into 
magnitudes using the UVOT photometric zero points \citep{2011AIPC.1358..373B}. 
The UVOT photometry is reported in Table~\ref{tab:uvot_mags} and presented in Figure~\ref{multiwave_lc}.

Several other instruments join the effort for the follow-up observations of GRB 190829A. 
We complemented the UVOT results including also the optical and near IR results obtained during the follow-up campaign and reported via Gamma-ray Coordinates Network (GCN) notices. We present in Figure~\ref{multiwave_lc} the IR measurements obtained by by GROND \citep{2019GCN.25569....1C}, UKIRT \citep{2019GCN.25584....1P} and TNG \citep{2019GCN.25591....1D} together with the optical fluxes measured by NOT \citep{2019GCN.25563....1H}, MMT \citep{2019GCN.25583....1F} and the Liverpool Telescope \citep{2019GCN.25585....1P,2019GCN.25592....1B}.     

\begin{table}
\begin{center}
 	\caption{UVOT photometry}
 	\label{tab:uvot_mags}
 	\begin{tabular}{lccc}
\hline
Time & Exposure & Filter & Afterglow \\
 (s) & (s) & & (AB mag) \\
\hline    
181  & 149.8  & $ white $   &  $> 21.8$   \\
556  & 19.8  & $ white $   & $21.0^{+1.3}_{-0.6}$   \\
731  & 19.8  & $ white $   & $19.9^{+0.4}_{-0.3} $  \\
945  & 149.8  & $ white $   & $18.20^{+0.06}_{-0.06} $  \\
1270  & 192.9  & $ white $   & $17.39^{+0.07}_{-0.06} $ \\
1610  & 182.4  & $ white $   & $17.59^{+0.09}_{-0.09} $   \\
6179  & 199.8  & $ white $   & $19.14^{+0.08}_{-0.08}  $ \\
11178  & 906.9  & $ white $   & $20.09^{+0.07}_{-0.07}$  \\
5.68 d  & 97551  & $ white $   & $23.3^{+1.0}_{-0.5}$  \\
10.73 d  & 166966  & $ white $   & $23.2^{+0.8}_{-0.5}$   \\
14.11 d  & 189777  & $ white $   & $24.0^{+2.9}_{-0.7}$    \\
17.10 d  & 177765  & $ white $   & $23.5^{+1.6}_{-0.6}$   \\
607  & 19.8  & $ v $   & $18.34^{+0.8}_{-0.5} $    \\
781  & 19.8  & $ v $   & $17.6^{+0.4}_{-0.3} $   \\
1147  & 192.5  & $ v $   & $15.9^{+0.1}_{-0.1}$  \\
1493  & 193.1  & $ v $   & $15.61^{+0.09}_{-0.09} $ \\
5154  & 199.8  & $ v $   & $17.2^{+0.1}_{-0.1} $ \\
6591  & 199.7  & $ v $   & $17.7^{+0.1}_{-0.1} $ \\
18062  & 906.8  & $ v $   & $19.7^{+0.3}_{-0.2} $ \\
45487  & 907.1  & $ v $   & $20.7^{+0.9}_{-0.5} $   \\
10.10 d  & 172695  & $ v $   & $21.8^{+3.5}_{-0.7} $  \\
12.35 d  & 206452  & $ v $   & $21.2^{+1.1}_{-0.5} $  \\
531  & 19.8  & $ b $   & $19.4^{+1.0}_{-0.5} $  \\
706  & 19.8  & $ b $   & $19.4^{+1.0}_{-0.5} $   \\
1245  & 193.5  & $ b $   & $17.0^{+0.1}_{-0.1}$   \\
1591  & 192.4  & $ b $   & $17.1^{+0.1}_{-0.1}$   \\
5974  & 199.8  & $ b $   & $18.6^{+0.1}_{-0.1}$   \\
7375  & 128.6  & $ b $   & $19.1^{+0.4}_{-0.3}$  \\
389  & 249.8  & $ u $   & $22.0^{+1.6}_{-0.6} $  \\
682  & 19.8  & $ u $   & $20.6^{+2.4}_{-0.7}  $ \\
855  & 19.8  & $ u $   & $19.4^{+0.6}_{-0.4}  $ \\
1221  & 193.8  & $ u $   & $18.0^{+0.2}_{-0.1}$   \\
1566  & 192.7  & $ u $   & $18.4^{+0.2}_{-0.2} $  \\
5769  & 199.8  & $ u $   & $19.4^{+0.2}_{-0.1} $  \\
7205  & 199.8  & $ u $   & $19.7^{+0.3}_{-0.2} $ \\
1.20 d  & 11080  & $ u $   & $22.3^{+1.5}_{-0.6}$   \\
657  & 19.8  & $ uvw1 $   & $19.9^{+1.0}_{-0.5}$    \\
830  & 19.8  & $ uvw1 $   & $20.4^{+1.9}_{-0.7}$   \\
1196  & 193.9  & $ uvw1 $   & $19.4^{+0.5}_{-0.3}$   \\
5564  & 199.8  & $ uvw1 $   & $20.7^{+0.4}_{-0.3} $  \\
7000  & 199.8  & $ uvw1 $   & $21.1^{+0.8}_{-0.5} $  \\
24108  & 899.8  & $ uvw1 $   & $21.7^{+0.7}_{-0.4} $\\
633  & 19.7  & $ uvm2 $   & $>$ 19.5 \\
1468  & 193.2  & $ uvw2 $   & $> 19.4$    \\
4950  & 199.8  & $ uvw2 $   & $> 20.9$  \\
6386  & 199.8  & $ uvw2 $   & $> 20.8$  \\
    \hline
    \end{tabular}
    \begin{flushleft}
    \quad \footnotesize{Time from the GRB trigger is expressed in seconds for observations taken the first day. Otherwise, it is given in days, as denoted with a d. Values are corrected for galactic extinction}\\
    \end{flushleft}
\end{center}
\end{table}

\begin{table}
\begin{center}
 	\caption{Stars used for calibration of Master photometry}
 	\label{tab:master-calib}
 	\begin{tabular}{lccc}
\hline
Gaia DR2 ID & RA & DEC &  $g$ AB mag\\
\hline
5179234683527590000  &  02:58:03.142  &  -08:56:15.40  &  16.01 \\
5179234889686020000  &  02:58:05.698  &  -08:55:11.03  &  13.33 \\
5179236813831370000  &  02:58:32.549  &  -08:56:08.77  &  16.09 \\
5179236813831370000  &  02:58:34.442  &  -08:55:45.12  &  15.54 \\
5179237741544310000  &  02:58:21.624  &  -08:54:42.44  &  15.58 \\
5179237844623520000  &  02:58:25.397  &  -08:54:38.66  &  15.57 \\
5179241040079190000  &  02:58:11.554  &  -08:53:03.08  &  14.81 \\
5179241040079190000  &  02:58:10.958  &  -08:52:25.97  &  15.30 \\
    \hline
    \end{tabular}
    \begin{flushleft}
    \end{flushleft}
\end{center}
\end{table}

\subsection{MASTER polarization measurements}

MASTER is a wide-field  fully robotic 40cm telescopes with identical scientific equipment at every observatory \citep[][]{Lipunov2010,Lipunov2019,Kornilov2012} and own real-time auto-detection system at 8: MASTER-Amur, -Tunka, -Ural, -Kislovodsk, -Tavrida (Russia), -SAAO (South Africa), -IAC (Spain, Tenerife), -OAFA (Argentina, Sah Juan University observatory OAFA). This equipment includes photometer with BVRI and 2 perpendicular oriented polarization filters oriented to 45~deg at neighboring MASTER observatories, that give us possibility to detect linear polarization \citep[][]{Troja2017,Gorbovskoy2016,Sadovnichy2018} or to observe the object 
simultaneously.
The MASTER network promptly reacted to the \textit{Swift} trigger and attempted to observe the XRT localization region \citep{Evans2019GCN} with the Kislovodsk telescope about 15 s after the BAT trigger.
Unfortunately the target was below the horizon at this time, and
the first images were obtained only 1239 s after trigger. 
The GRB optical counterpart was detected by the MASTER auto-detection system \cite[][]{Lipunov2010,Lipunov2019} at 3.1$\arcsec$ east and
7.5$\arcsec$ south from the galaxy center.

The source was observed with the clear ($C$) and the polarization filters ($P$). The clear band magnitude $C$ is best described by the Gaia $g$ filter. 
Polarized observations were taken with broadband polarizers manufactured using linear conducting nanostructure technology \citep[][]{Kornilov2012}. 
Magnitudes obtained from broadband $P$ filters correspond to 0.2$B$ + 0.8$R$ where $B$ and $R$ are the standard Johnson filters.
For MASTER-Kislovodsk, the orientation of the polarizing filters was 0$^\circ$ and 90$^\circ$, counted counter-clockwise from the north direction.
Photometry was calibrated using 8 nearby stars from the catalog Gaia DR2 (Table~\ref{tab:master-calib}). 

In addition to these reference stars, we selected a large list of comparison stars with brightness similar to the afterglow. This allowed us to determine the magnitude error as function of the luminosity variation of stars (see \citealt{Lipunov2019} and \citealt{Troja2017} for a more detailed description).
After astrometric binding of each image, we performed standard aperture photometry using an aperture size of 3.9$\arcsec$.
To assess the galaxy's contribution, we used MASTER archived images taken about one and a half year after the GRB detection (starting at 19:18:19 UT of March 14, 2021). 
The surface brightness of the host galaxy was estimated to 22.2 mag/arcsec$^2$ and subtracted from the afterglow light curve. 

\begin{table*}
\begin{center}
 	\caption{Master photometry}
 	\label{tab:master}
 	\resizebox{.63\textwidth}{!}{%
 	\begin{tabular}{lcccccc}
\hline
Time$^{a}$ & Exposure & Filter &  Instrument & WEST$^{b}$ & EAST$^{c}$ & $|Q|$\\
    s & s & & & ABmag & ABmag & \\
    \hline
 ..  &  5   &  $C$   &  MASTER-Amur        &   $> 14.2$          &        -           &     -          \\
 ..  &  35   &  $C$   &  MASTER-Amur        &   $> 13.9$          &        -           &     -          \\
 ..  &  70   &  $C$   &  MASTER-Amur        &   $> 13.9$          &        -           &     -          \\
1758 &  180  &  $P$   &  MASTER-Kislovodsk  &   $14.6 \pm 0.2$  &   $14.3 \pm 0.1$ & $0.14 \pm 0.09$ \\
1988 &  180  &  $P$   &  MASTER-Kislovodsk  &   $14.4 \pm 0.2$  &   $14.1 \pm 0.2$ & $0.12 \pm 0.12$ \\
2440 &  180  &  $P$   &  MASTER-Kislovodsk  &   $14.5 \pm 0.2$  &   $14.44 \pm 0.07$ & $0.01 \pm 0.09$ \\
2650 &  180  &  $P$   &  MASTER-Kislovodsk  &   $14.43 \pm 0.07$  &   $14.5 \pm 0.2$ & $0.03 \pm 0.08$ \\
2881 &  180  &  $P$   &  MASTER-Kislovodsk  &   $15.0 \pm 0.1$  &   $14.7 \pm 0.2$ & $0.13 \pm 0.12$ \\
3085 &  180  &  $P$   &  MASTER-Kislovodsk  &   $15.10 \pm 0.09$  &   $14.8 \pm 0.2$ & $0.16 \pm 0.11$ \\
3316 &  180  &  $P$   &  MASTER-Kislovodsk  &   $14.88 \pm 0.06$  &        $14.95 \pm 0.1^{d}$ & $0.03 \pm 0.05^{d}$          \\
3521 &  180  &  $P$   &  MASTER-Kislovodsk  &   $15.13 \pm 0.08$  &   $15.11 \pm 0.07$ & $0.01 \pm 0.05$ \\
3746 &  180  &  $P$   &  MASTER-Kislovodsk  &   $15.19 \pm 0.08$  &   $15.20 \pm 0.08$ & $0.00 \pm 0.05$ \\
3950 &  180  &  $P$   &  MASTER-Kislovodsk  &   $15.23 \pm 0.08$  &   $15.17 \pm 0.08$ & $0.03 \pm 0.05$ \\
4181 &  180  &  $P$   &  MASTER-Kislovodsk  &   $15.34 \pm 0.09$  &   $15.41 \pm 0.09$ & $0.03 \pm 0.06$ \\
4386 &  180  &  $P$   &  MASTER-Kislovodsk  &   $15.6 \pm 0.1$  &   $15.6 \pm 0.1$ & $0.01 \pm 0.06$ \\
4617 &  180  &  $P$   &  MASTER-Kislovodsk  &   $15.8 \pm 0.1$  &   $15.7 \pm 0.1$ & $0.03 \pm 0.07$ \\
4835 &  180  &  $P$   &  MASTER-Kislovodsk  &   $16.1 \pm 0.1$  &        $15.8 \pm 0.1^{d}$ & $0.11 \pm 0.07^{d}$        \\
5070 &  180  &  $P$   &  MASTER-Kislovodsk  &   $16.0 \pm 0.1$  &   $15.9 \pm 0.1$ & $0.03 \pm 0.08$ \\
5274 &  180  &  $P$   &  MASTER-Kislovodsk  &   $16.1 \pm 0.1$  &   $15.9 \pm 0.1$ & $0.08 \pm 0.08$ \\
5505 &  180  &  $P$   &  MASTER-Kislovodsk  &   $15.99 \pm 0.13$  &   $16.05 \pm 0.13$ & $0.03 \pm 0.08$ \\
5714 &  180  &  $P$   &  MASTER-Kislovodsk  &   $16.0 \pm 0.1$  &   $16.1 \pm 0.1$ & $0.04 \pm 0.09$ \\
5948 &  180  &  $P$   &  MASTER-Kislovodsk  &   $16.3 \pm 0.2$  &   $16.3 \pm 0.2$ & $0.01 \pm 0.10$ \\
    \hline
    \end{tabular}
    }
    \begin{flushleft}
    \quad \footnotesize{Values are corrected for galactic extinction}\\
    \quad \footnotesize{$^{a}$ Seconds since the BAT trigger}\\
    \quad \footnotesize{$^{b}$ Magnitudes obtained with the polarimeter mounted on the WEST tube}\\
    \quad \footnotesize{$^{c}$ Magnitudes obtained with the polarimeter mounted on EAST tube}\\
    \quad \footnotesize{$^{d}$ The EAST tube magnitude at this specific time was obtained by interpolating the light curve}\\
    \end{flushleft}
\end{center}
\end{table*}

\begin{table*}
\begin{center}
 	\caption{ATCA and ALMA afterglow observations}
 	\label{tab:atca}
 	\resizebox{.90\textwidth}{!}{%
 	\begin{tabular}{lcccccccccc}
\hline
Time$^{a}$ & Date & Start Obs & End Obs & $F_{2.1 GHz}$ & $F_{5.5 GHz}$ & $F_{9.0 GHz}^{b}$ & $F_{16.7 GHz}$ & $F_{21.2 GHz}$ & $F_{99.8~GHz}^{c}$ & $F_{343.5~GHz}$\\
Days & & UT & UT & mJy & mJy & mJy & mJy & mJy & mJy & mJy\\
    \hline
0.84     &  2019-08-30  &  14:49  &  17:11  &  $0.78 \pm 0.05$     &  $2.05 \pm 0.07$   &  $3.2 \pm 0.1$     &        ---        &   --- &   --- &   --- \\ 
1.48     &  2019-08-31  &  6:59  &  8:09  &  ---     &  ---   &  ---     &   ---   &   ---    &  ---   &  $9.9 \pm 1.0$ \\
1.56     &  2019-08-31  &  9:20 &  9:34   &  ---     &  ---   &  ---     &   ---   &   ---    &  $7.3 \pm 0.4$   &  --- \\    
2.07     &  2019-08-31  &  20:34  &  22:30  &  $1.38 \pm 0.07$     &  $3.3 \pm 0.1$     &  $4.0 \pm 0.1$     &  $4.1 \pm 0.2$    &  $4.0 \pm 0.2$ &   --- &   --- \\
2.45     &  2019-09-01  &  6:42 &  6:56   &  ---     &  ---   &  ---     &   ---   &   ---    &  $6.0 \pm 0.3$   &  --- \\
2.48     &  2019-09-01  &  7:09  &  7:36  &  ---     &  ---   &  ---     &   ---   &   ---    &  $6.0 \pm 0.3$   &  --- \\
2.80     &  2019-09-01  &  14:55  &  15:32  &         ---          &  $3.2 \pm 0.1$     &  $3.3 \pm 0.1$     &        ---        &   --- &   --- &   --- \\ 
3.48     &  2019-09-02  &  7:22 &  7:36  &  ---     &  ---   &  ---     &   ---   &   ---   &  $4.6 \pm 0.2$ &  --- \\
4.04     &  2019-09-02  &  20:00  &  22:02  &  $3.5 \pm 0.1$       &  $4.0 \pm 0.1$     &  $3.1 \pm 0.1$     &  $2.7 \pm 0.1$    &  $2.8 \pm 0.1$ &   --- &   --- \\
5.49     &  2019-09-04  &  7:47 &  8:01  &  ---     &  ---   &  ---     &   ---   &   ---  &  $2.6 \pm 0.1$ &  --- \\
7.05     &  2019-09-05  &  20:07  &  22:05  &  $1.09 \pm 0.07$     &  $2.07 \pm 0.08$   &  $2.47 \pm 0.09$   &  $2.6 \pm 0.2$    &  $1.07 \pm 0.08$ &   --- &   --- \\
8.99     &  2019-09-07  &  19:34  &  19:59  &         ---          &  $1.77 \pm 0.07$   &  $2.40 \pm 0.08$   &        ---        &   --- &   --- &   --- \\ 
12.00    &  2019-09-10  &  19:51  &  20:13  &         ---          &         ---        &  $1.08 \pm 0.05$   &        ---        &   --- &   --- &   --- \\ 
14.05    &  2019-09-12  &  20:11  &  22:02  &  $0.85 \pm 0.06$     &  $1.34 \pm 0.08$   &         ---        &  $1.19 \pm 0.07$  &  $1.03 \pm 0.07$ &   --- &   --- \\
15.06    &  2019-09-13  &  21:04  &  21:35  &         ---          &  $1.00 \pm 0.06$   &         ---        &        ---        &   --- &   --- &   --- \\ 
18.48    &  2019-09-17  &  7:24 &  7:38  &  ---     &  ---   &  ---     &   ---   &   ---   &  $0.38 \pm 0.04$ &  --- \\
20.37    &  2019-09-19  &  4:40  &  5:09  &  ---     &  ---   &  ---     &   ---   &   ---    &  $0.24 \pm 0.03$   &  --- \\
21.41    &  2019-09-20  &  5:15  &  6:23  &  ---     &  ---   &  ---     &   ---   &   ---    &  ---   &  $0.16 \pm 0.03$ \\
23.89    &  2019-09-22  &  16:02  &  18:30  &      $<0.3 $        &  $0.69 \pm 0.05$   &  $0.67 \pm 0.04$   &  $0.31 \pm 0.05$  &  $<0.20$ &   --- &   --- \\
24.58    &  2019-09-23  &  9:53 &  10:07  &  ---     &  ---   &  ---     &   ---   &   ---  &  $0.25 \pm 0.04$ &  --- \\
27.33    &  2019-09-26  &  3:57 &  4:11  &  ---     &  ---   &  ---     &   ---   &   ---   &  $0.23 \pm 0.03$ &  --- \\
30.37    &  2019-09-29  &  4:05  &  5:43  &  ---     &  ---   &  ---     &   ---   &   ---    &  $0.26 \pm 0.03$   &  $<0.10$ \\
31.36    &  2019-09-30  &  4:36 &  4:50  &  ---     &  ---   &  ---     &   ---   &   ---     &  $0.19 \pm 0.02$ &  ---   \\
31.81    &  2019-09-30  &  15:00  &  15:54  &         ---          &  $0.40 \pm 0.04$   &  $0.44 \pm 0.04$   &        ---        &  --- &   --- &   --- \\ 
38.63    &  2019-10-07  &  10:33  &  11:29  &         ---          &  $0.23 \pm 0.07$   &  $0.30 \pm 0.06$   &        ---        &  --- &   --- &   --- \\ 
44.98    &  2019-10-13  &  19:07  &  19:56  &         ---          &  $0.33 \pm 0.07$   &  $0.20 \pm 0.06$   &        ---        &  --- &   --- &   --- \\ 
87.90    &  2019-11-25  &  14:07  &  21:01  &      $<0.1$         &  $0.18 \pm 0.03$   &      $<0.06$       &        ---        &  --- &   --- &   --- \\
    \hline
    \end{tabular}
    }
    \begin{flushleft}
    \quad \footnotesize{Upper limits are given at $3\sigma$ level}\\
    \quad \footnotesize{$^{a}$ Days since the BAT trigger}\\
    \quad \footnotesize{$^{b}$ The central frequency is 9.5 GHz instead of 9.0 GHz for September 7 and September 10 observations}\\
    \quad \footnotesize{$^{c}$ The central frequency is 97.5 GHz instead of 99.8 GHz for August 31 at 6:59 UT, September 19 and September 29 observations}\\
    \end{flushleft}
\end{center}
\end{table*}

\subsection{ALMA and ATCA radio observations}

Data were collected from the Atacama Large Millimetre-Submillimetere Array (ALMA) science archive under programs 2018.1.01410.T (PI: Perley) and 2018.1.01454.T (PI: Laskar). 
Observations were carried out in Band 3 and Band 7, at 99.8 GHz and 335 GHz central frequencies respectively, at different epochs corresponding to Sept 1, Sept 2, Sept 4, Sept 17, Sept 19, Sept 23, Sept 26, Sept 29 and Sept 30 for Band 3, and Aug 31, Sept 20 and Sept 29 for Band 7. 
For our analysis we used the final calibrated images retrieved from the archive.
Photometric measurements of RMS imaged noise and Gaussian fits were performed within CASA (Common AStronomy Software Applications; version 5.4; \citealt{McMullin2007}) utilizing the
\textsc{viewer()} function tools. The same regions were selected in all the data sets, with only reliable converged fits reported. The angular resolution of ALMA for these observations was roughly 0.230-0.31$\arcsec$ for Band 3 and 0.67-0.89$\arcsec$ for Band 7. 
Results are summarized in Table~\ref{tab:atca}. Flux density error bars were estimated summing in quadrature the rms measured from final calibrated images and the calibration ALMA uncertainty for data collected in band 3 and band 7 (which correspond to 5\% and 10\% of the flux, respectively \footnote{https://almascience.nrao.edu/documents-and-tools/cycle8/alma-proposers-guide}).

Data were collected from the Australia Telescope Compact Array (ATCA) radio telescope under programs C3289 (PI: Laskar) and  C3316 (PI: Lekshmi). 
We processed the public dataset through the data reduction package {\it Miriad} \citep{Sault1995} using standard procedures for the 
5.5, 9.0, 9.5, 16.7 and 21.2~GHz datasets, which included splitting, manual flagging, calibration and clean \& restore imaging. The 2.1 GHz datasets were split out and flagged via the sumthreshold algorithm \citep{Offringa2010} implemented in the {\it Miriad} task {\it pgflag} using the user guide suggested settings. The C3289 datasets use the source $0238-084$ as phase calibrator. The C3316 datasets use $0237-027$ as a phase calibrator, instead.
All datasets use the source $1934-638$ as bandpass calibrator and for the flux density scale bootstrap.
The ATCA observations until 2019 Sep 13th were taken 6C extended array configuration and thus imaged with robustness parameter robust=0 and all antennas were included. 
Observations past 2019 Sep 13th were performed in H168 compact hybrid array configuration; they were imaged with robustness parameter robust=2 and antenna-6 visibilities were excluded. Phase decorrelation, more evident in the 16.7-21.2~GHz datasets was accounted for and fixed by comparing triple product flux outputs from the {\it Miriad} task {\it calred} with restored image peak flux density values in phase calibrator scans. Flux density error bars were estimated by summing in quadrature the multiplicative (systematic) error (related to complex gain calibration residuals) with the rms noise measured in V-Stokes maps. The multiplicative term was estimated at 3\% at 2.1, 5.5, 9.0 and 9.5~GHz, and 5\% at 16.7 and 21.2~GHz. Flux densities and $3\sigma$ upper limits for the 14 epochs analysed are presented in Table~\ref{tab:atca}.

\section{Results}\label{sec:results}

\subsection{Prompt emission}
The gamma-ray emission is characterized by two distinct episodes with different spectral properties. For the first short ($\sim$~5 s) peak, BAT measures a photon index~$\sim 1.9$, softer than the majority of BAT GRB spectra \citep{Lien2016}. This value is consistent with the location of the peak energy close to the energy bandpass  ($E_{peak}^{obs}$=$130 \pm 20$~keV), as derived from the {\it Fermi}/GBM analysis \citep{2019GCN.25575....1L, Chand20}. The isotropic-equivalent energy ($E_{\gamma,iso}$) of the first pulse is $2.6 \times 10^{49}$ erg. As noted by \citet{Chand20}, these values are inconsistent with the correlation between the rest frame peak energy ($E_{peak}^{src}$) and the isotropic radiated energy ($E_{\gamma,iso}$) observed for other long GRBs \citep{Amati2002}. 

\begin{figure}
\centering
\includegraphics[scale=0.22, clip]{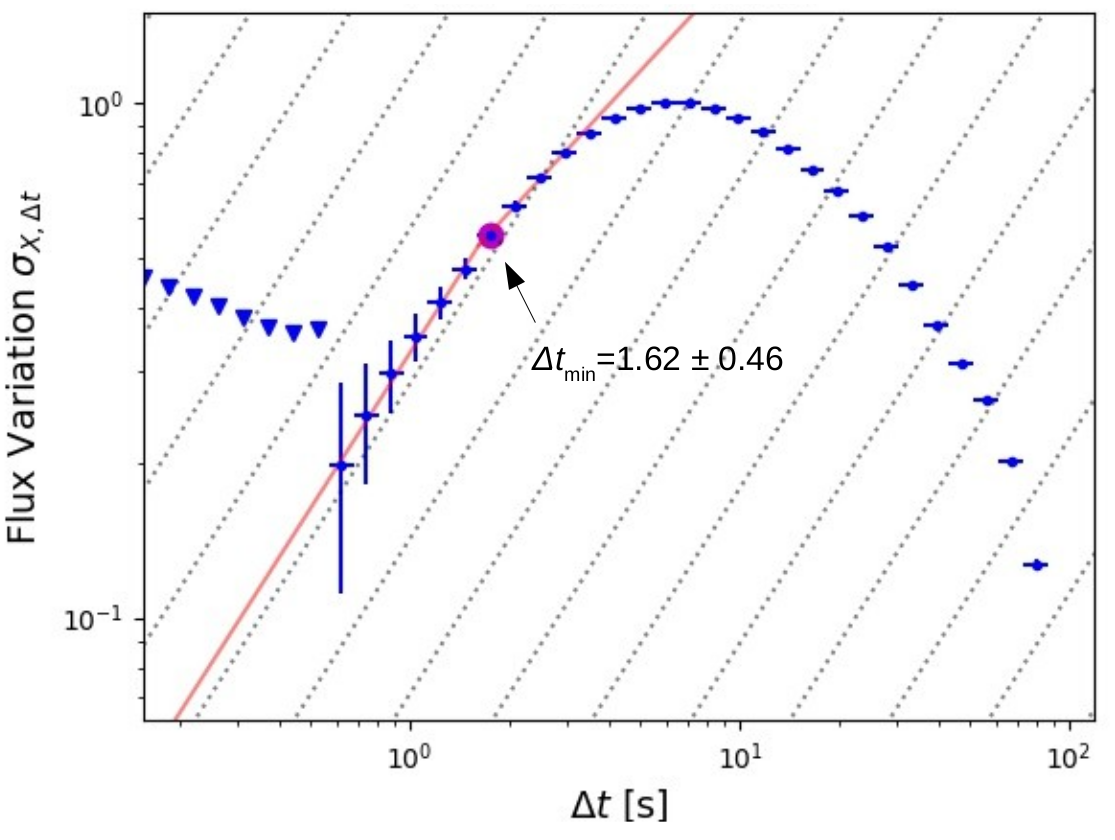}
\caption{Haar wavelet scaleogram $\sigma_{X,\Delta~t}$ vs. observer frame timescale $\Delta~t$ obtained following the same procedure presented by \citep{Golkhou2014} The minimum variability timescale is identified by the point in which the scalogram (red line) deviate from the straight line parallel to the dotted lines and it is reported on top of the plot (observer frame). 
\label{min_timescale}}
\end{figure}

On the other side, the spectrum of the second episode (the main peak in Fig.~\ref{fig:LCs}) 
is even softer with a photon index $\Gamma$\,$\approx$2.6 at the extreme of the distribution observed by \textit{Swift} \citep{Lien2016}. According to the definition presented in \cite{sakamoto2008}, this very soft burst can be classified as an X-ray flash (XRF),
similar to other nearby events such as GRB~100316D \citep{Starling2011} and GRB~060218 \citep{Soderberg06} which belong to the X-ray rich (XRR) GRBs class. 
By describing the prompt spectrum with a canonical Band function \citep{band1993} of low-energy index $\alpha$=1.0 and high-energy index $\beta$=2.3, 
the soft photon index of this burst implies a low peak energy, at the edge or below that BAT energy bandpass.  Using the Band model, we derive an upper limit to the peak energy of $E_{peak}^{obs} <$ 24 keV (90\% confidence level) and a fluence of $f_{\gamma} = (5.76 \pm 0.10) \times 10^{-6}~\mathrm{erg\,cm^{-2}}$ (15--150 keV). 
\begin{figure*}
\centering
\includegraphics[scale=0.90]{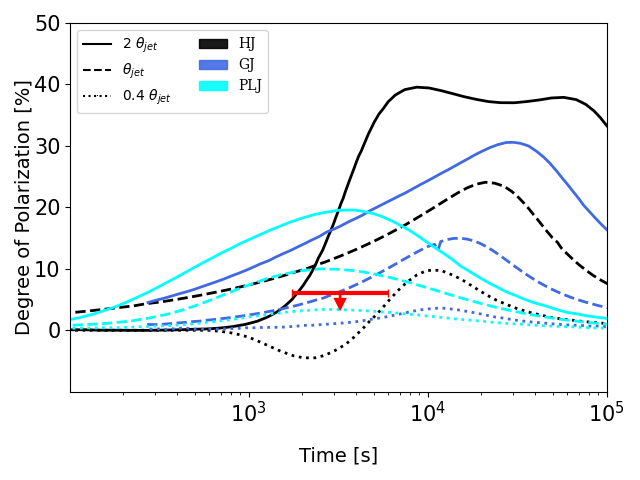}
\caption{
MASTER optical upper limit on linear polarization compared with different theoretical models \citep[from ][]{Rossi2004} 
for off-axis angles of 2$\theta_{jet}$, $\theta_{jet}$ and 0.4 $\theta_{jet}$ (solid, dashed and dotted line, respectively). 
Black, blue and cyan lines refer to a homogeneous (HJ), Gaussian (GJ) and power-law (PLJ) jet structure, respectively. The red downward arrow show the polarization limit obtained for favorable orientations of the polarization angle (e.g. $\Theta=0$).
\label{fig:polarizarion_models}
}
\end{figure*}
The isotropic equivalent energy released from the main pulse is $E_{iso} \lesssim 2 \times 10^{50}$ erg (1-10000 keV), which makes the second pulse of GRB~190829A consistent with the Amati correlation for long GRBs \citep[see also][]{Chand20}.

In order to better characterize the gamma-ray signal, we also studied the temporal properties of the prompt emission, the shape of the light curve,  its variability and the time delay between the the precursor and the main peak. 
The temporal profile of the gamma-ray emission visibly differs from the ones observed in other low-redshift bursts. For the other three nearby GRBs detected by {\it Swift}, namely GRB~060218, GRB100316D, and GRB171205A,  the light curve displays a broad and rather smooth profile with a duration of two hundreds seconds or longer. 
For GRB 190829A the main peak follows a fast rise with exponential decay (FRED; \citealt{Norris96}) profile with an unremarkable duration of $\approx$15~s, more similar to the prompt phase observed by {\it Beppo/SAX} for GRB~980425 \citep{Pian2000}. 
Moreover, none of the other nearby bursts have a precursor similar to the one observed for this burst.

 The quiescent time between the first and the second peak is about 40 seconds. 
 This delay is substantially longer than the average quiescent time measured in other GRBs observed by {\it Swift} \citep[about 20; s][]{Burlon2008}, falling in the top 30\% of measured delays between the precursor and the main peak. 
 However, the sample of GRBs with precursors studied in \citet{Burlon2008} was rather small in size and, when comparing GRB~190829A to the larger sample of bursts observed by {\it Fermi} \citep{Coppin2020},  its quiescent time is higher than average but still within 1~$\sigma$ of the observed distribution \citep[see Figure 3 in ][]{Coppin2020}.
 We therefore conclude that the long delay between precursor and main peak is not an uncommon property among other GRBs. 

Another interesting parameter used to characterize the prompt GRB emission is its minimum variability timescale, which we derived using the tecnique of \citet{Golkhou2014} based on the Haar wavelets. 

From this analysis, we found $\Delta~t_{min}/(1+z)=1.5 \pm 0.5$ (see Figure~\ref{min_timescale}). This timescale is higher than the median value obtained for {\it Swift} GRBs \citep[$\sim$0.5 s in the rest-frame; ][]{Golkhou2014}. This slow variability combined with the soft spectra and low gamma-ray luminosity agrees with previous studies of the prompt emission that found correlation between these parameters \citep[e.g. ][]{Dichiara2016,Rizzuto2007,Guidorzi2005,Reichart2001,Fenimore2000}.    
It also indicate that a slow component with a soft spectrum is dominating the signal. 
This could be ascribed to viewing angle effects \citep{Sato2021}, thermal emission from the photosphere \citep{Vetere2006,ZhangICMART} or the cocoon break-out. 

\subsection{Early afterglow}

\subsubsection{Polarization}\label{sec:polar}

MASTER did not detect any optical polarization 
at the time of the afterglow peak, observing between $T_0$+1700~s and $T_0$+6000~s (Figure~\ref{multiwave_lc}). In this interval, the time-averaged  3~$\sigma$ upper limit
is $Q$\,$\lesssim$6\%, where $Q=\left(F_{0}-F_{90}\right)/\left(F_0 + F_{90}\right)$ is the Stokes parameter,
$F_{0}$ and $F_{90}$ are the fluxes measured by the two perpendicular filters.
We use this limit to constrain the presence of an off-axis jet. 

Synchrotron radiation from a collimated relativistic outflow is 
expected to show a substantial degree of polarization when observed off-axis \citep[e.g.,][]{Rossi2004,GillGranot2018,GillGranot2020}. 
As the jet decelerates, the emitting surface becomes asymmetrical for an observer
misaligned with respect to the jet's axis. 
Therefore, the linear polarization  of the observed emission increases in time and
reaches its peak when $\Gamma$\,$\approx$\,$(\theta_v - \theta_j)^{-1}$, where 
$\Gamma$ is the Lorentz factor, 
$\theta_v$ the viewing angle, and $\theta_j$ the jet's characteristic angular width. 
This roughly corresponds to the peak of the afterglow light curve. 
The degree of polarization depends on a variety of factors, most importantly the 
configuration of the magnetic field. 
Values of polarization as high as 40\% are expected if the magnetic field lies along the shock plane, whereas the presence of a parallel $B$ component would decrease this estimate. 

In the case of GRB~190829A, the delayed X-ray and optical peak was interpreted by \cite{Sato2021} as emission from a narrow jet ($\theta_j$\,$\lesssim$1$^{\circ}$) observed
at an off-axis angle $\theta_{v}~\sim~2\theta_{jet}$. 
Assuming the same jet parameters of \citet{Sato2021}, we compare our polarization limit with the predictions 
of off-axis emission using the different jet structures described in \cite{Rossi2004}: 
a homogeneous (or top-hat), a Gaussian, and a power-law angular profile.  
In Figure~\ref{fig:polarizarion_models} we show the values derived assuming a
magnetic field along the plane of the shock. In all cases with $\theta_{v}~\gtrsim~\theta_{jet}$ our limit is well below
the predictions for off-axis jet emission. 

However, each telescope of the MASTER network has two filter with orthogonal polarization angles and can only constrain the component of the polarization vector along the line of sight. 
The Stokes parameter $Q$ is related to the total degree of linear polarization $P_L$ as
$Q$\,=$P_L$ cos(2 $\Theta$) where $\Theta$ is the polarization angle. 
For unfavorable orientations, that is $\Theta$\,$\gtrsim$\,1/2 arccos($Q_{UL}$/$P_L$), a highly polarized afterglow could still result in a non-detection \citep{Gorbovskoy2012}. 
For a randomly oriented polarization vector, the probability of a non detection is however small and we can disfavour the presence of an off-axis jet with $\theta_{v}~\sim~2\theta_{jet}$ with a confidence of $\approx$90\% for a homogeneous jet, 
$\approx$80\% for a power-law jet, and $\approx$70\% for a Gaussian jet.

\begin{figure}
\centering
\includegraphics[scale=0.455, clip]{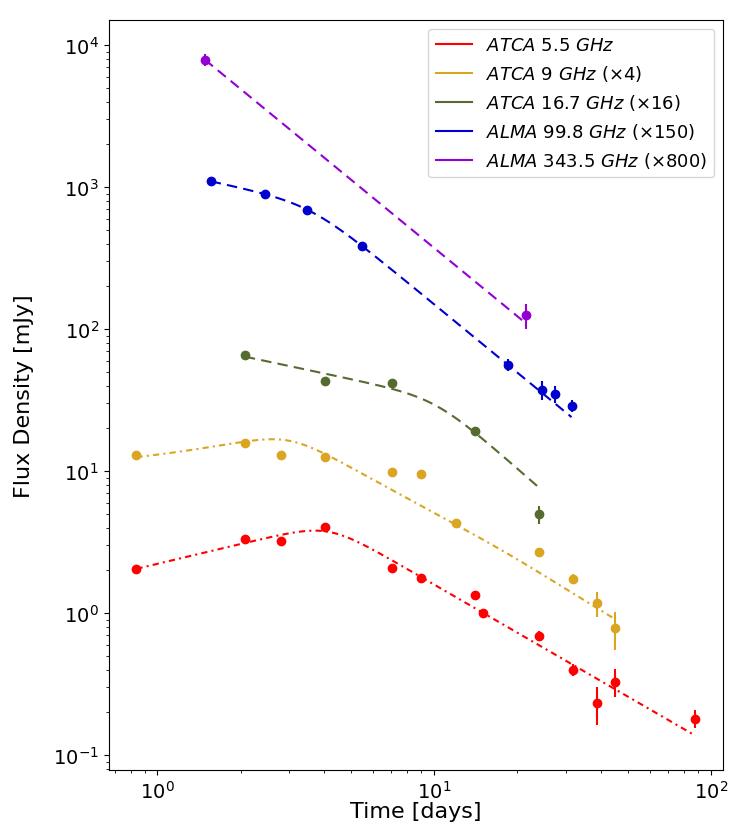}
\caption{Radio light curves at different frequencies. Fluxes are re-scaled for clarity.
Dash-dotted and dashed curve show the curves dominated by the forward shock and the reverse shock component, respectively. 
\label{radio_lc}}
\end{figure}
\begin{figure}
\centering
\includegraphics[scale=0.35, clip]{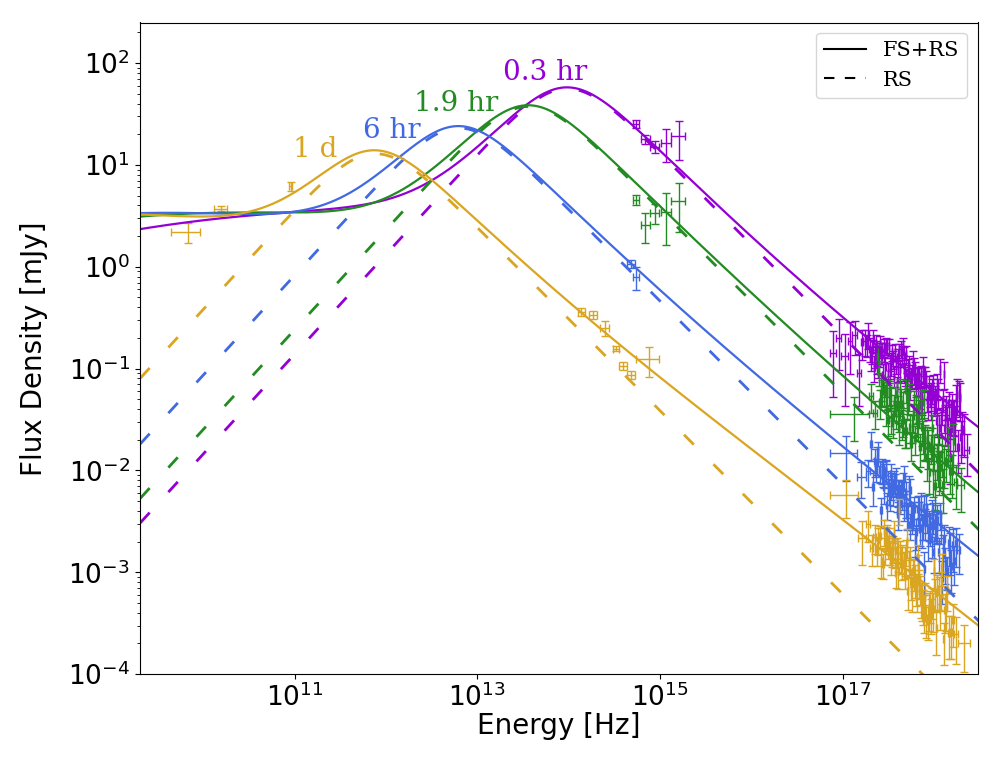}
\caption{Early afterglow SEDs at 4 different epochs:
0.3 hr, 1.9 hr, 6 hr and 1 d. 
For each SED, the solid line shows the best fit model consisting of a FS and a RS component, 
the dashed line shows only the RS contribution. 
\label{fig:sed_plot}}
\end{figure}

\subsubsection{A forward shock plus reverse shock scenario}\label{FRshock}

The multi-wavelength afterglow of GRB~190829A presents a complex evolution, as shown in Figure~\ref{multiwave_lc}. 
The X-ray light curve shows a initial steep decay, $F_X$ $\propto$ $t^{-\alpha}$ with index $\alpha$\,$\sim$2.5 up to $\sim$ 180 s, followed by a first small peak at about 400~s and a prominent rebrightening starting at $\sim$700 s. 
By modeling the light curve with a smoothly broken power-law (dashed line in Figure~\ref{multiwave_lc}), we derive that the emission rises with a slope of $-3.2 \pm 0.5$, reaches its peak at around 1400~s and  then decays with index $1.65 \pm 0.18$. 
The decay becomes shallower after $\approx$ $2.3 \times 10^4$~s when the index changes to $\alpha_X$\,$\approx$1.0. A second temporal break appears at $\approx 3 \times 10^5$ s 
with a steepening to $\alpha_X$ = 1.26 $\pm$0.06. 

The bright hump at 1400~s  is also observed in the UV, optical and nIR light curves. 
From the fit with a smoothly broken power-law, we derive a steep rising index
$\alpha_{\rm UVOIR}$ = -3.1 $\pm$ 0.4, a peak time of $\approx$1500~s, and a
decay index $\alpha_{\rm UVOIR}$ = 1.71 $\pm$ 0.18.
The data also indicate a temporal break at 
$\approx 2-3 \times 10^{4}$ s when the afterglow transitions to a shallower decay 
with index  $\alpha_{\rm UVOIR}$\,$\approx$1.1.   

The achromatic nature of the late peak suggests that the UV/optical/nIR and X-ray emission 
 arise from the same spectral component. 
 The lack of significant polarization disfavors models of off-axis jets and viewing angle effects to explain the delayed afterglow peak (Section~\ref{sec:polar}). 
 The sharp temporal rise, $\alpha \sim$-3,  and the steep post-peak decay, $\alpha$\,$\sim$1.7, are not consistent with the standard evolution of external shock emission, and the achromatic nature of the rebrightening is not typical of X-ray flares either.
 The most likely explanation is that the observed optical and X-ray rebrightening is dominated by the reverse shock component, identified by \citet{Rhodes2020} in the late-time radio light curves. 

Our analysis of the radio data largely confirms the results of \citet{Rhodes2020}. 
At the highest frequency probed by ALMA (343 GHz), 
the afterglow rapidly decays as $t^{-1.65}$. 
At lower frequencies, between 100~GHz and 17~GHz, a broken power-law, 
with temporal indices $\approx$0.45 and $\approx$1.65, is required to describe the data.  
The break times are 3.6$\pm$0.1 days and 10.1$\pm$2.9 days for 100~GHz and 17~GHz, respectively. These values are consistent with the evolution expected from a reverse shock that is expanding in a constant density medium. Furthermore, they can not be explained by a standard forward shock model \citep[e.g. ][]{Zhang2021} because it significantly underestimate the emission observed at high radio frequencies (see Appendix~\ref{sec:appendix} for more details).
The steep decay after the break is similar to the post-peak decay observed in the optical and X-rays afterglow (also with index $\sim$1.65). This supports the hypothesis of a common origin for the broadband emission: a bright RS emission initially peaking 
close to the optical/nIR range, then shifting toward radio wavelengths as it cools off. 
The shallow decay of the radio flux tracks the peak of the reverse shock. 
As also noted by \citet{Rhodes2020},  the measured temporal decays are shallower than model predictions, and may indicate a temporal evolution of the afterglow parameters or continued energy injection. 

At even lower frequencies ($<$9 GHz), a different behavior is observed. 
The radio flux at 5.5 GHz increases with a slope of -0.46 $\pm$ 0.09 reaching its peak at 4.1 $\pm$ 0.4 days, and then decays with slope of 1.13 $\pm$ 0.12 (Figure~\ref{radio_lc}). 
As suggested by \citet{Rhodes2020}, the low-frequency behavior is consistent with emission
from a forward shock expanding into a homogeneous medium. 

From the well-sampled radio light curve we can extract basic afterglow parameters, such as the peak flux $F_p$\,$\approx$\,4 mJy and synchrotron frequency $\nu_m$\,$\approx$\,5.5~GHz at 4 days. Using the standard closure relations for GRB afterglows, the post peak temporal slope, $\alpha \approx$1.1,  is related to the spectral index of the accelerated electrons 
as $p$ = 4$\alpha$/3 +1 = 2.50 $\pm$0.16.
This is softer, although consistent within the 2~$\sigma$ interval, than the value
inferred from spectroscopy of the VHE component \citep{HESS2021} and may indicate that the electrons' energy distribution is not described by a simple power-law but presents a harder component at high energies.

\paragraph{Spectral Energy Distribution (SED)}

To better disentangle the different emission components we studied the afterglow SED at four different epochs: 1200~s (slightly before the peak), 6800~s (during the steep decay), 21600~s (around the time of the H.E.S.S. detection),  and 1~d after the BAT trigger. 
Each component was described with a smoothly broken power-law (SBPL), and its evolution was determined according to the standard fireball model.  The first spectral component tracks the evolution of a forward shock (FS) expanding in a constant density medium, 
as observed at the low frequency radio data. 
The peak frequency $\nu_{m,FS}$ is located between the nIR and the radio data, and evolves as $t^{-3/2}$. 
Assuming an electron spectral energy index $p$\,$\sim$\,2.50, as derived from the radio light curves, we set 
$\beta^{FS}_{low}$=-1/3 and $\beta^{FS}_{up}$=0.75 for the FS spectral index below and above the peak energy, respectively.
The smoothness parameter $s$, which determines the shape of the spectral breaks \citep{Granot2002}, was left free to vary. 

\begin{figure}
\includegraphics[scale=0.48]{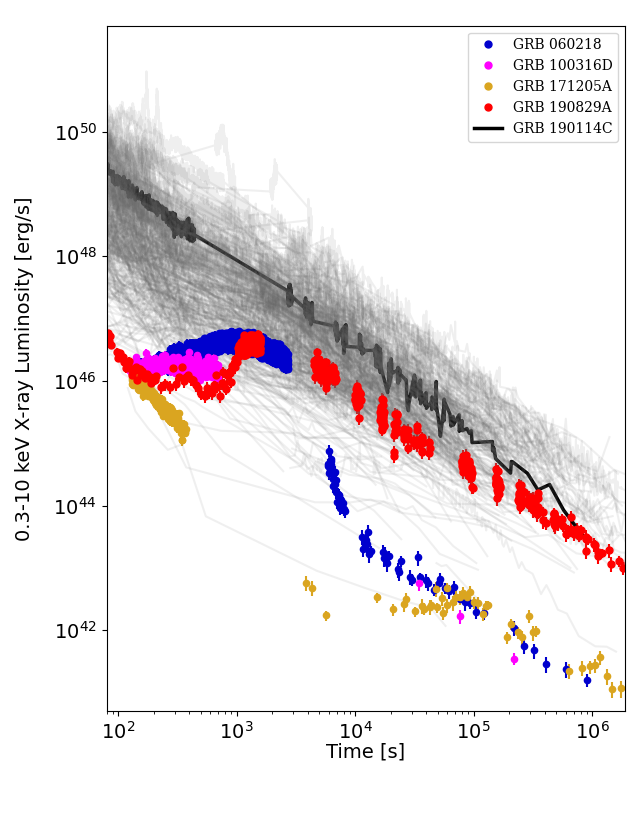}
\includegraphics[scale=0.48]{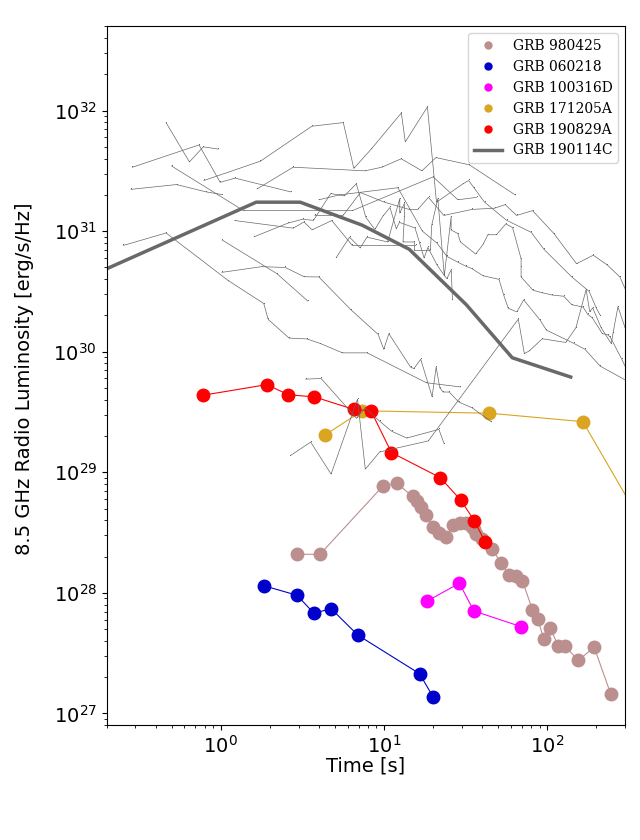}
\caption{Top: X-ray luminosity light curves (0.3-10 keV; rest frame) of GRB 190829A (red) compared with the sample of  nearby long GRBs detected by {\it Swift}. 
The gray curves show the range of luminosities measured for a sample of 321 long GRBs with known redshift. 
The other burst with VHE emission, GRB~190114C \citep{MAGIC2019}, is highlighted. 
Bottom: Radio luminosity light curve (8.5 GHz; rest-frame) 
of GRB~190829A (red) compared with the sample of  nearby long GRBs. 
The gray curves shows the range of luminosities measured for a sample of long GRBs with known redshift (from \citealt{ChandraFrail2012})
\label{fig:X-ray_lum}.
}
\end{figure}

The second spectral component tracks the reverse shock (RS) modeled with spectral slopes of $\beta^{RS}_{low}$\,$\approx$\,-1.0 $\pm$0.2 and $\beta^{RS}_{up}$\,$\approx$\,0.90$\pm$0.02, leading to $p$=2.80$\pm$0.04. 
The RS spectral peak is located at $\approx$\,9$\times$ 10$^{13}$ Hz at the first epoch ($T_{0}$+1200 s) and decays approximately 
as  $\nu_{m,RS} \propto t^{-3/2}$. 
By imposing the standard closure relations for a RS expanding in a constant density medium, no acceptable fit could be found. 
In particular, the RS peak frequency ($\nu_{m,RS}$) is higher than the one observed for the FS ($\nu_{m,FS}$) and  the RS peak flux was found to decrease in time with a power-law index of $\sim$-0.4, whereas a value of $\sim$-1.0 is expected for a thick shell RS. This behaviour is different than the one expected from the standard theory \citep{Kobayashi2000,Zhang2003}.
The observed peak frequency ratio ($\nu_{m,RS}$/$\nu_{m,FS}$) could be explained by a particularly low initial Lorentz factor (e.g. $\Gamma_{0}\sim$10) and a high ratio between the magnetic field enclosed in the RS compared with the one in the FS ($B_{RS}$/$B_{FS}\sim$ 200).

The best fit SEDs are shown in Figure~\ref{fig:sed_plot}. 
They are obtained for an intrinsic reddening of $E(B-V)$=$0.90 \pm 0.05$ and an intrinsic hydrogen column desnity of $N_H=1.13 \pm 0.06 \times 10^{22}$ cm$^{-2}$, 
consistent with the analysis of \citet{Chand20}. 
This indicates an extremely dusty environment, also found in GRB~190114C.  
Our model shows a dominant RS component in the UV/optical/nIR range up to $\approx$1~d after the burst. The underlying FS component begins to emerge several hours after the burst, 
and might account, at least partially, for the blue excess visible in the  optical spectra at 0.3~d \citep{Hu2021}.

\subsubsection{Burst Energetics and Environment}

GRBs in the local universe are generally followed by faint afterglows produced by sub-energetic explosions. 
With a total gamma-ray energy of $\approx$\,2$\times 10^{50}$ erg, GRB~190829A lies at the intersection between classical GRBs and sub-luminous nearby events. 

As shown in Figure~\ref{fig:X-ray_lum} (top panel), three local events (060218, 100316D, and 171205A) were all characterized by low-luminosity X-ray afterglows with $L_X \approx 10^{43}$\,erg\,s$^{-1}$ at 1 d, orders of magnitude below the average luminosity of cosmological long GRBs. 
At late times, the X-ray band is likely independent from the circumburst density, and directly probes the total energy of the explosion. 

GRB~190829A does not cluster around other nearby events, but is characterized by a long-lived and luminous X-ray afterglow. Its behavior  tracks the general trend of cosmological GRBs, although it occupies the lower end of the observed luminosity distribution. 
We use its X-ray luminosity at 14~d, $L_X$\,$\approx 1.7 \times 10^{43}$\,erg\,s$^{-1}$, to estimate the total isotropic-equivalent blastwave energy \citep{Granot2002}:
\begin{equation}
\label{eq1}
E_{K,\rm iso} \approx 3 \times 10^{52} \left(\frac{\epsilon_{e}}{0.1}\right)^{-\frac{4}{3}} 
\left(\frac{\epsilon_{B}}{10^{-4}}\right)^{-\frac{1}{9}} {\rm erg}
\end{equation}
where we use $z$=0.00785, $\nu_X$\,$\sim$5 keV, and assume $p$=2.5. We further assume that the fraction of emitting electrons is $\xi$=1. 
From the study of the prompt emission we derived a $\gamma$-ray isotropic-equivalent energy of $\approx 2 \times 10^{50}$ erg and a low radiative efficiency  $\eta_{\gamma}$\,$\lesssim$\,1\%.

Our estimated total energy is an order of magnitude lower than the value derived by \cite{Salafia2021}, however we caution that our value can only place a lower limit to the burst energetics. 
If $\xi$\,$<$1, as suggested by simulations \citep{Sironi2011}, the lower fraction of emitting electrons would require a larger energy budget to reproduce the observed X-ray luminosity.
Furthermore, the presence of a VHE component in GRB~190829A suggests that Compton losses might be significant and strongly suppress the X-ray flux \citep{Beniamini2015}. In this case, the kinetic energy would be higher by a factor $\approx$\,$(1+Y)$, where $Y$ is the Compton parameter.

At radio wavelengths, the GRB afterglow initially appears brighter than the radio counterparts of local bursts.
This is likely the consequence of the strong RS component crossing the radio range. 
At later times, the GRB radio luminosity is instead comparable with the sample of nearby sub-energetic events. 
The shallow temporal decay with $\alpha$\,$\approx$1.1 excludes that the low radio luminosity is due to a jet-break. 
As the FS peak crosses the radio band a few days after the burst, the radio afterglow is in the slow cooling regime with 
$\nu_{sa}<\nu_m<\nu_r<\nu_c$ and is sensitive to the density of the surrounding medium, $F_R \propto n^{1/2}$ \citep{Granot2002}. 
Its faintness may therefore be a consequence of a low density environment,
which is also consistent with a low self-absorption frequency 
$\nu_{sa}\lesssim$1~ GHz. 
By using the total energy in Equation~\ref{eq1} and the ratio 
log $(E_K/n) \approx$ 54 from \citet{Salafia2021}, we infer a circumburst density $n\approx$0.01 - 0.1\,cm$^{-3}$, typical of the interstellar medium. 
The relatively low density of the circumburst region might indicate that the high intrinsic absorption measured through X-ray spectroscopy, $N_H$=$1.13 \times 10^{22}$ cm$^{-2}$,
is not local to the explosion site, but due to a distant intervening absorber. 

Finally, we can use our observations to set a lower limit to the jet-break time $t_j\gtrsim100$~d. For the energies and density estimated above, this corresponds to a jet opening angle \citep{Rhoads99,Sari99}:

\begin{equation}
\theta_j \gtrsim 16^{\circ}  \left(\frac{t_j}{100~d}\right)^{\frac{3}{8}} 
\left(\frac{\epsilon_{e}}{0.1}\right)^{\frac{1}{6}} 
\left(\frac{n}{0.1 {\rm~cm}^{-3}}\right)^{\frac{1}{8}} 
\end{equation}

and a collimation-corrected energy release $E_K \gtrsim 5 \times 10^{50} {\rm~erg}$.
This value is consistent with the general population of GRBs \citep[e.g.][]{Aksulu21} and suggests that, unlike other nearby sub-energetic bursts, GRB~190829A was analogue to the class of cosmological GRBs. 
\\
~

\section{Summary and Conclusions}\label{sec:conclusions}

We analyzed the early afterglow of GRB 190829A studying multi-wavelength observations that include {\it Swift} X-ray, optical and UV data, optical polarization measurements obtained by the MASTER global network, and radio data obtained by ALMA and ATCA. 

The lack of detectable optical polarization allows us to disfavor emission from a misaligned jet with $\theta_{v}\,\gtrsim\,2\theta_{jet}$ \citep[e.g.][]{Sato2021}, indicating that viewing angle effects can not explain the bright and delayed peak observed at optical and X-ray wavelengths. 
Our study of the multi-color light curves and broadband SEDs favors a forward shock plus reverse shock model \citep{Rhodes2020,Salafia2021} in which the early afterglow peak is powered by the reverse shock emission. 
From this model, we derive a spectral index $p\approx2.5$ for the electrons' energy distribution, softer than the value derived from the study of the VHE emission ($p\approx$2.1; \citealt{HESS2021}). 

Although the forward/reverse shock model naturally describes the salient features of the afterglow, its basic version, as described by the standard fireball model and closure relations \citep[][]{vanderHorst2014}, 
does not provide a good fit to the dataset.
This can be due to different factors: continued energy injection, time-dependent shock parameters, or a non trivial density profile \citep[e.g.][]{Gao2013,Sanchez-Ramirez2017,Ryan2020}. In the case of GRB~190829A, the prompt gamma-ray emission shows an energetic precursor preceding the main event by $\approx$40~s, and the interaction between the precursor fireball and the main peak fireball could lead to a non-canonical evolution of the early afterglow. 
In this scenario, discussed by \citet{Nappo2014}, the precursor and the main event are produced by the same central engine. A first standard afterglow is produced by the precursor fireball and its interaction with the circumburst medium. After a period of quiescence, characterized by a low (but non-zero) injection rate, the second fireball is emitted by the main prompt event. This interacts with the earlier shells until it catches up with the precursor fireball, producing a second afterglow component. The properties of the two fireballs and their afterglows may substantially differ, generating a non-standard afterglow evolution. 

Finally, we conclude that GRB~190829A lies in the intermediate region between classical GRBs and
nearby sub-energetic events, adding an important element to understand the origin of engine-driven explosions. 
We place a lower limit to the total energy budget $\gtrsim 5 \times 10^{50} {\rm~erg}$, 
which lies in the typical range of cosmological long GRBs. 
We also find a wide jet opening angle $\approx$16~deg and a low radiative efficiency $<$1\%, indicative
of a weak prompt gamma-ray emission. 
The latter could be explained if the wide jet is viewed only slightly off-axis ($\theta_v<\theta_j$) or
if the emerging jet was in part stalled by the stellar envelope.

\section*{ACKNOWLEDGEMENTS}
This work was initiated during the “Astrophysics in the LIGO/Virgo Era” meeting at the Aspen Center for Physics during Summer 2019. The Aspen Center for Physics is supported by National Science Foundation grant PHY-1607611."
MASTER is supported by Lomonosov Moscow State University Development program. VL, EG, VK were supported by RFBR 19-29-11011 grant.
This paper makes use of the following ALMA data: ADS/JAO.ALMA\#2018.1.01410.T,  ADS/JAO.ALMA\#2018.1.01454.T. ALMA is a partnership of ESO (representing its member states), NSF (USA) and NINS (Japan), together with NRC (Canada), MOST and ASIAA (Taiwan), and KASI (Republic of Korea), in cooperation with the Republic of Chile. The Joint ALMA Observatory is operated by ESO, AUI/NRAO and NAOJ. The National Radio Astronomy Observatory is a facility of the National Science Foundation operated under cooperative agreement by Associated Universities, Inc.
The work was also partially supported by the National Aeronautics and Space Administration through grant 80NSSC18K0429 issued through the Astrophysics Data Analysis Program.

\section*{Data Availability}
The data underlying this article will be shared on reasonable request to the corresponding author.


\appendix

\section{Standard Forward Shock model}
\label{sec:appendix}

\begin{figure*}
\centering
\includegraphics[scale=0.70]{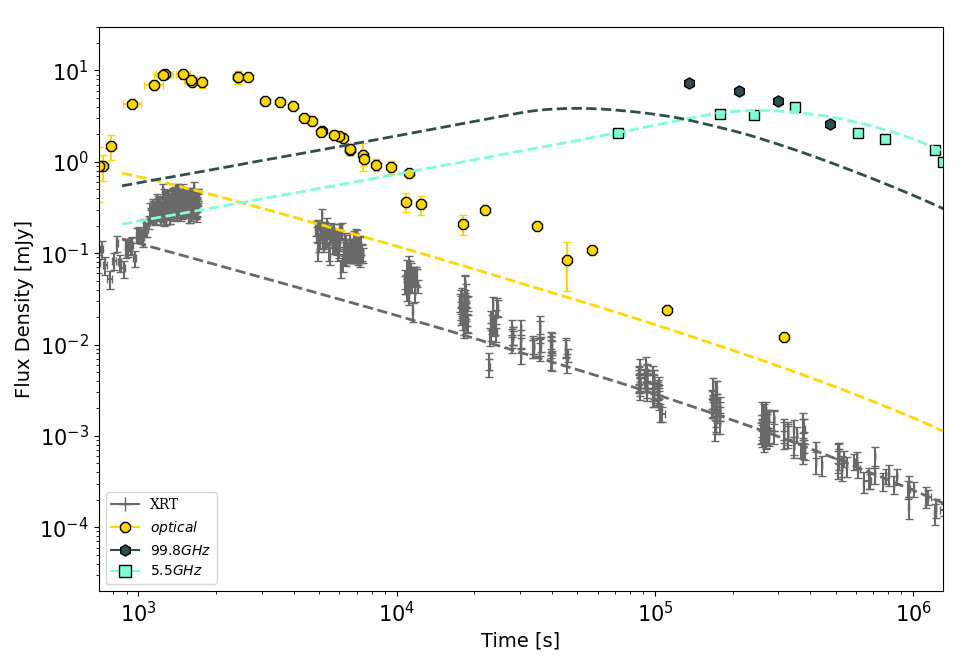}\\
\caption{
Dashed lines show the best fit model prediction for a simple forward shock model expanding in a constant density profile. While the late X-ray data and the low-frequency radio light curve are consistent with the model, the early afterglow is substantially underestimated as well as the high-frequency radio data points
\label{fig:appendix}.
}
\end{figure*}

We used \texttt{afterglowpy}\footnote{https://github.com/geoffryan/afterglowpy} \citep{Ryan2020} to describe the multi-wavelength data with the predictions of a standard FS propagating in a constant density medium. We assume a gaussian jet structure observed on-axis. The results are shown in Figure~\ref{fig:appendix}
for the following parameters: $E_K$\,$\approx$\,$8 \times 10^{51}$ erg, $\theta_j$\,$\approx$\,$25^{\circ}$, a fraction of accelerated electrons $\xi_N$\,$\approx$\,$0.35$, $n$\,$\approx$\,$10$ $cm^{-3}$, 
$p$\,$\approx$\,$2$, $\epsilon_{e}$\,$\approx$\,$2\times 10^{-2}$ and $\epsilon_{B}$\,$\approx$\,$3 \times 10^{-5}$.  \\
While the late X-rays and low-frequency radio data can be reasonably well described by this FS model, the early time X-rays/optical data and the high-frequency radio emission are significantly underestimated by the model. We interpret this as evidence of an additional component of emission, consistent with a bright RS  (as discussed in Section~\ref{FRshock}).



\label{lastpage}
\end{document}